# A Regularity Lemma, and Low-weight Approximators, for Low-degree Polynomial Threshold Functions

Ilias Diakonikolas[*]   Rocco A. Servedio[†]   Li-Yang Tan[‡]

Andrew Wan[§]

Department of Computer Science
Columbia University
{ilias, rocco, liyang, atw12}@cs.columbia.edu

October 25, 2018

## Abstract

We give a "regularity lemma" for degree-$d$ polynomial threshold functions (PTFs) over the Boolean cube $\{-1, 1\}^n$. Roughly speaking, this result shows that every degree-$d$ PTF can be decomposed into a constant number of subfunctions such that almost all of the subfunctions are close to being regular PTFs. Here a "regular" PTF is a PTF $\text{sign}(p(x))$ where the influence of each variable on the polynomial $p(x)$ is a small fraction of the total influence of $p$.

As an application of this regularity lemma, we prove that for any constants $d \geq 1, \epsilon > 0$, every degree-$d$ PTF over $n$ variables can be approximated to accuracy $\epsilon$ by a constant-degree PTF that has integer weights of total magnitude $O(n^d)$. This weight bound is shown to be optimal up to logarithmic factors.

[*]Supported by NSF grant CCF-0728736, and by an Alexander S. Onassis Foundation Fellowship. Part of this work was done while visiting IBM Almaden.

[†]Supported by NSF grants CCF-0347282, CCF-0523664 and CNS-0716245, and by DARPA award HR0011-08-1-0069.

[‡]Supported by DARPA award no. HR0011-08-1-0069 and NSF CyberTrust grant no. CNS-0716245.

[§]Supported by NSF CyberTrust award CNS-0716245.

# 1 Introduction

A polynomial threshold function (henceforth PTF) is a Boolean function $f : \{-1,1\}^n \to \{-1,1\}$, $f(x) = \text{sign}(p(x))$, where $p : \{-1,1\}^n \to \mathbb{R}$ is a polynomial with real coefficients. If $p$ has degree $d$, we say that $f$ is a *degree-$d$* PTF. Low-degree PTFs are a natural generalization of linear threshold functions (the case $d = 1$) and hence are of significant interest in complexity theory, see e.g. [ABFR94, Bru90, OS03b, OS03a, DRST09, GL94, HKM09, MZ09, Sak93, She09] and many other works.

The *influence* of coordinate $i$ on a function $g : \{-1,1\}^n \to \mathbb{R}$ measures the extent to which $x_i$ affects the output of $g$. More precisely, we have $\text{Inf}_i(g) = \sum_{S \ni i} \widehat{g}(S)^2$, where $\sum_{S \subseteq [n]} \widehat{g}(S) \chi_S(x)$ is the Fourier expansion of $g$. The *total influence* of $g$ is the sum of all $n$ coordinate influences, $\text{Inf}(g) = \sum_{i=1}^n \text{Inf}_i(g)$. See [O'D07a, KS06] for background on influences.

We say that a polynomial $p : \{-1,1\}^n \to \mathbb{R}$ is "$\tau$-regular" if the influence of each coordinate on $p$ is at most a $\tau$ fraction of $p$'s total influence (see Section 2 for a more detailed definition). A PTF $f$ is said to be $\tau$-regular if $f = \text{sign}(p)$, where $p$ is $\tau$-regular. Roughly speaking, regular polynomials and PTFs are useful because they inherit some nice properties of PTFs and polynomials over Gaussian (rather than Boolean) inputs; this intuition can be made precise using the "invariance principle" of Mossel et al. [MOO05]. This point of view has been useful in the $d = 1$ case for constructing pseudorandom generators [DGJ+09], low-weight approximators [Ser07, DS09], and other results for LTFs [OS08, MORS09].

## 1.1 Our results

**A regularity lemma for degree-$d$ PTFs.** A number of useful results in different areas, loosely referred to as "regularity lemmas," show that for various types of combinatorial objects an arbitrary object can be approximately decomposed into a constant number of "pseudorandom" sub-objects. The best-known example of such a result is Szemerédi's classical regularity lemma for graphs [Sze78], which (roughly) says that any graph can be decomposed into a constant number of subsets such that almost every pair of subsets induces a "pseudorandom" bipartite graph. Another example is Green's recent regularity lemma for Boolean functions [Gre05]. Results of this sort are useful because different properties of interest are sometimes easier to establish for pseudorandom objects, and via regularity lemmas it may be possible to prove the corresponding theorems for general objects. We note also that results of this sort play an important part in the "structure versus randomness" paradigm that has been prominent in recent work in combinatorics and number theory, see e.g. [Tao07].

We prove a structural result about degree-$d$ PTFs which follows the above pattern; we thus refer to it as a "regularity lemma for degree-$d$ PTFs." Our result says that any low-degree PTF can be decomposed as a small depth decision tree, most of whose leaves are close to regular PTFs:

**Theorem 1.** *Let $f(x) = \text{sign}(p(x))$ be any degree-$d$ PTF. Fix any $\tau > 0$. Then $f$ is equivalent to a decision tree $\mathcal{T}$, of depth*

$$\text{depth}(d, \tau) := \frac{1}{\tau} \cdot \big(d \log \frac{1}{\tau}\big)^{O(d)}$$

*with variables at the internal nodes and a degree-$d$ PTF $f_\rho = \text{sign}(p_\rho)$ at each leaf $\rho$, with the following property: with probability at least $1 - \tau$, a random path[1] from the root reaches a leaf $\rho$ such that $f_\rho$ is $\tau$-close to some $\tau$-regular degree-$d$ PTF.*

Regularity is a natural way to capture the notion of pseudorandomness for PTFs, and results of interest can be easier to establish for regular PTFs than for arbitrary PTFs (this is the case for our main application, constructing low-weight approximators, as we describe below). Our regularity lemma provides a general tool to reduce questions about arbitrary PTFs to regular PTFs; it has already been used in this way as an essential ingredient in the recent proof that bounded independence fools all degree-2 PTFs [DKN09]. We note that the recent construction of pseudorandom generators for degree-$d$ PTFs of [MZ09] also crucially uses a decomposition result which is very similar to our regularity lemma; we discuss the relation between our work and [MZ09] in more detail below and in Appendix C.

---

[1] A random path corresponds to the standard uniform random walk on the tree.

**Application: Every low-degree PTF has a low-weight approximator.** [Ser07] showed that every linear threshold function (LTF) over $\{-1, 1\}^n$ can be $\epsilon$-approximated by an LTF with integer weights $w_1, \ldots, w_n$ such that $\sum_i w_i^2 = n \cdot 2^{\tilde{O}(1/\epsilon^2)}$. (Here and throughout the paper we say that $g$ is an $\epsilon$-approximator for $f$ if $f$ and $g$ differ on at most $\epsilon 2^n$ inputs from $\{-1, 1\}^n$.) This result and the tools used in its proof found several subsequent applications in complexity theory and learning theory, see e.g. [DGJ+09, OS08].

We apply our regularity lemma for degree-$d$ PTFs to prove an analogue of the [Ser07] result for low-degree polynomial threshold functions. Our result implies that for any constants $d, \epsilon$, any degree-$d$ PTF has an $\epsilon$-approximating PTF of constant degree and (integer) weight $O(n^d)$.

When we refer to the *weight* of a PTF $f = \text{sign}(p(x))$, we assume that all the coefficients of $p$ are integers; by "weight" we mean the sum of the squares of $p$'s coefficients. We prove

**Theorem 2.** *Let $f(x) = \text{sign}(p(x))$ be any degree-$d$ PTF. Fix any $\epsilon > 0$. Then there is a polynomial $q(x)$ of degree $D = (d/\epsilon)^{O(d)}$ and weight $2^{(d/\epsilon)^{O(d)}} \cdot n^d$ such that $\text{sign}(q(x))$ is $\epsilon$-close to $f$.*

A result on the existence of low-weight $\epsilon$-approximators for PTFs is implicit in the recent work [DRST09]. They show that any degree-$d$ PTF $f$ has Fourier concentration $\sum_{|S|>1/\epsilon^{O(d)}} \widehat{f}(S)^2 \leq \epsilon$, and this easily implies that $f$ can be $\epsilon$-approximated by a PTF with integer weights. (Indeed, recall that the learning algorithm of [LMN93] works by constructing such a PTF as its hypothesis.) The above Fourier concentration bound implies that there is a PTF of degree $1/\epsilon^{O(d)}$ and weight $n^{1/\epsilon^{O(d)}}$ which $\epsilon$-approximates $f$. In contrast, our Theorem 2 can give a weaker degree bound (if $d = 1/\epsilon^{\omega(1)}$), but always gives a much stronger weight bound in terms of the dependence on $n$. We mention here that Podolskii [Pod09] has shown that for every constant $d \geq 2$, there is a degree-$d$ PTF for which any *exact* representation requires weight $n^{\Omega(n^d)}$.

We also prove lower bounds showing that weight $\widetilde{\Omega}(n^d)$ is required to $\epsilon$-approximate degree-$d$ PTFs for sufficiently small constant $\epsilon$; see Section 3.3.

**Techniques.** An important ingredient in our proof of Theorem 1 is a case analysis based on the "critical index" of a degree-$d$ polynomial (see Section 2 for a formal definition). The critical index measures the first point (going through the variables from most to least influential) at which the influences "become small;" it is a natural generalization of the definition of the "critical index" of a linear form [Ser07] that has been useful in several subsequent works [OS08, DGJ+09, DS09]. Roughly speaking we show that

- If the critical index of $p$ is large, then a random restriction fixing few variables (the variables with largest influence in $p$) causes $\text{sign}(p)$ to become a close-to-constant function with non-negligible probability; see Section 2.1. (Note that a constant function is trivially a regular PTF.)
- If the critical index of $p$ is positive but small, then a random restriction as described above causes $p$ to become regular with non-negligible probability; see Section 2.2.
- If the critical index of $p$ is zero, then $p$ is already a regular polynomial as desired.

**Related Work.** The results of Sections 2.1 and 2.2 strengthen earlier results with a similar flavor in [DRST09]. Those earlier results had quantitative bounds that depended on $n$ in various ways: getting rid of this dependence is essential for our low-weight approximator application and for the application in [DKN09].

Simultaneously and independently of this work, Ben-Eliezer et al. [BELY09], Harsha et al. [HKM09], and Meka and Zuckerman [MZ09] have proved similar structural results for PTFs. In particular, [HKM09] give a result which is very similar to Lemma 11, the main component in our proof of Theorem 1. By applying the result from [HKM09], Meka and Zuckerman [MZ09] give a result which is quite similar to our Theorem 1. However, their definition of regularity is somewhat different from ours, and as a consequence their structural results and ours are quantitatively incomparable, as we discuss in Appendix C. Ben-Eliezer et al. give a result of similar flavor as our Theorem 1. They establish the existence of a decision tree such that most leaves are $\tau$-regular (as opposed to $\tau$-close to being $\tau$-regular). The depth of their tree is exponential in $1/\tau$, which makes it quantitatively weaker for applications.

## 1.2 Preliminaries

We start by establishing some basic notation. We write $[n]$ to denote $\{1, 2, \ldots, n\}$ and $[k, \ell]$ to denote $\{k, k+1, \ldots, \ell\}$. We write $\mathbf{E}[X]$ and $\mathbf{Var}[X]$ to denote expectation and variance of a random variable $X$, where the underlying distribution will be clear from the context. For $x \in \{-1, 1\}^n$ and $A \subseteq [n]$ we write $x_A$ to denote $(x_i)_{i \in A}$.

We assume familiarity with the basic elements of Fourier analysis over $\{-1, 1\}^n$; a concise review of the necessary background is given in Appendix A.1. Let $p : \{-1, 1\}^n \to \mathbb{R}$ and $p(x) = \sum_S \widehat{p}(S)\chi_S(x)$ be its Fourier expansion. The *influence* of variable $i$ on $p$ is $\mathrm{Inf}_i(p) \stackrel{\text{def}}{=} \sum_{S \ni i} \widehat{p}(S)^2$, and the *total influence* of $p$ is $\mathrm{Inf}(p) = \sum_{i=1}^n \mathrm{Inf}_i(p)$. For a function $f : \{-1, 1\}^n \to \mathbb{R}$ and $q \geq 1$, we denote by $\|f\|_q$ its $l_q$ norm, i.e. $\|f\|_q \stackrel{\text{def}}{=} \mathbf{E}_x[|p(x)|^q]^{1/q}$, where the intended distribution over $x$ will always be uniform over $\{-1, 1\}^n$.

For Boolean functions $f, g : \{-1, 1\}^n \to \{-1, 1\}$ the distance between $f$ and $g$, denoted $\mathrm{dist}(f, g)$, is the $\mathbf{Pr}_x[f(x) \neq g(x)]$ where the probability is over uniform $x \in \{-1, 1\}^n$.

Our proofs will use various bounds from probability theory, which we collect for easy reference in Appendices A.2 and A.3. We call the reader's attention in particular to Theorem 7; throughout the paper, $C$ (which will be seen to play an important role in our proofs) denotes $C_0^2$, where $C_0$ is the universal constant from that theorem.

## 2 Main Result: a regularity lemma for low-degree PTFs

Let $f : \{-1, 1\}^n \to \{-1, 1\}$ be a degree-$d$ PTF. Fix a representation $f(x) = \mathrm{sign}(p(x))$, where $p : \{-1, 1\}^n \to \mathbb{R}$ is a degree-$d$ polynomial which (w.l.o.g.) we may take to have $\mathbf{Var}[p] = 1$. We assume w.l.o.g. that the variables are ordered in such a way that $\mathrm{Inf}_i(p) \geq \mathrm{Inf}_{i+1}(p)$ for all $i \in [n-1]$.

We now define the notion of the $\tau$-critical index of a polynomial [DRST09] and state its basic properties.

**Definition 1.** *Let $p : \{-1, 1\}^n \to \mathbb{R}$ and $\tau > 0$. Assume the variables are ordered such that $\mathrm{Inf}_j(p) \geq \mathrm{Inf}_{j+1}(p)$ for all $j \in [n-1]$. The $\tau$-critical index of $p$ is the least $i$ such that:*

$$\mathrm{Inf}_{i+1}(p) \leq \tau \cdot \sum_{j=i+1}^n \mathrm{Inf}_j(p). \tag{1}$$

*If (1) does not hold for any $i$ we say that the $\tau$-critical index of $p$ is $+\infty$. If $p$ has $\tau$-critical index $0$, we say that $p$ is $\tau$-regular.*

Note that if $p$ is a $\tau$-regular polynomial then $\max_i \mathrm{Inf}_i(p) \leq d\tau$ since the total influence of $p$ is at most $d$. If $f(x) = \mathrm{sign}(p(x))$, we say $f$ is $\tau$-regular when $p$ is $\tau$-regular, and we take the $\tau$-critical index of $f$ to be that of $p$. [2] The following lemma (see Appendix B for the easy proof) says that the total influence $\sum_{i=j+1}^n \mathrm{Inf}_i(p)$ goes down geometrically as a function of $j$ prior to the critical index:

**Lemma 1.** *Let $p : \{-1, 1\}^n \to \mathbb{R}$ and $\tau > 0$. Let $k$ be the $\tau$-critical index of $p$. For $j \in [0, k]$ we have*
$$\sum_{i=j+1}^n \mathrm{Inf}_i(p) \leq (1-\tau)^j \cdot \mathrm{Inf}(p).$$

---

[2] Strictly speaking, $\tau$-regularity is a property of a particular representation and not of a PTF $f$, which could have many different representations. The particular representation we are concerned with will always be clear from context.

We will use the fact that in expectation, the influence of an unrestricted variable in a polynomial does not change under random restrictions (again see Appendix B for the easy proof):

**Lemma 2.** *Let $p : \{-1, 1\}^n \to \mathbb{R}$. Consider a random assignment $\rho$ to the variables $x_1, \ldots, x_k$ and fix $\ell \in [k+1, n]$. Then $\mathbf{E}_\rho[\mathrm{Inf}_\ell(p_\rho)] = \mathrm{Inf}_\ell(p)$.*

**Notation:** For $\mathcal{S} \subseteq [n]$, we write "$\rho$ fixes $\mathcal{S}$" to indicate that $\rho \in \{-1, 1\}^{|\mathcal{S}|}$ is a restriction mapping $x_\mathcal{S}$, i.e. each coordinate in $\mathcal{S}$, to either $-1$ or $1$ and leaving coordinates not in $\mathcal{S}$ unrestricted.

## 2.1 The large critical index case

The main result of this section is Lemma 3, which says that if the critical index of $f$ is large, then a noticeable fraction of restrictions $\rho$ of the high-influence variables cause $f_\rho$ to become close to a constant function.

**Lemma 3.** *Let $f : \{-1, 1\}^n \to \{-1, 1\}$ be a degree-$d$ PTF $f = \mathrm{sign}(p)$. Fix $\beta > 0$ and suppose that $f$ has $\tau$-critical index at least $K \overset{def}{=} \alpha/\tau$, where $\alpha = \Omega(d \log \log(1/\beta) + d \log d)$. Then, for at least a $1/(2C^d)$ fraction of restrictions $\rho$ fixing $[K]$, the function $f_\rho$ is $\beta$-close to a constant function.*

*Proof.* Partition the coordinates into a "head" part $H \overset{def}{=} [K]$ (the high-influence coordinates) and a "tail" part $T = [n] \setminus H$. We can write $p(x) = p(x_H, x_T) = p'(x_H) + q(x_H, x_T)$, where $p'(x_H)$ is the truncation of $p$ comprising only the monomials all of whose variables are in $H$, i.e. $p'(x_H) = \sum_{S \subseteq H} \widehat{p}(S) \chi_S(x_H)$.

Now consider a restriction $\rho$ of $H$ and the corresponding polynomial $p_\rho(x_T) = p(\rho, x_T)$. It is clear that the constant term of this polynomial is exactly $p'(\rho)$. To prove the lemma, we will show that for at least a $1/(2C^d)$ fraction of all $\rho \in \{-1, 1\}^K$, the (restricted) degree-$d$ PTF $f_\rho(x_T) = \mathrm{sign}(p_\rho(x_T))$ satisfies $\mathbf{Pr}_{x_T}[f_\rho(x_T) \neq \mathrm{sign}(p'(\rho))] \leq \beta$. Let us define the notion of a *good restriction*:

**Definition 2.** *A restriction $\rho \in \{-1, 1\}^K$ that fixes $H$ is called* good *iff the following two conditions are simultaneously satisfied: (i) $|p'(\rho)| \geq t^* \overset{def}{=} 1/(2C^d)$, and (ii) $\|q(\rho, x_T)\|_2 \leq t^* \cdot \left(\Theta(\log(1/\beta))\right)^{-d/2}$.*

Intuitively condition (i) says that the constant term $p'(\rho)$ of $p_\rho$ has "large" magnitude, while condition (ii) says that the polynomial $q(\rho, x_T)$ has "small" $l_2$-norm. We claim that if $\rho$ is a good restriction then the degree-$d$ PTF $f_\rho$ satisfies $\mathbf{Pr}_{x_T}[f_\rho(x_T) \neq \mathrm{sign}(p'(\rho))] \leq \beta$. To see this claim, note that for any fixed $\rho$ we have $f_\rho(x_T) \neq \mathrm{sign}(p'(\rho))$ only if $|q(\rho, x_T)| \geq |p'(\rho)|$, so to show this claim it suffices to show that if $\rho$ is a good restriction then $\mathbf{Pr}_{x_T}[|q(\rho, x_T)| \geq |p'(\rho)|] \leq \beta$. But for $\rho$ a good restriction, by conditions (i) and (ii) we have

$$\mathbf{Pr}_{x_T}[|q(\rho, x_T)| \geq |p'(\rho)|] \leq \mathbf{Pr}_{x_T}\left[|q(\rho, x_T)| \geq \|q(\rho, x_T)\|_2 \cdot \left(\Theta(\log(1/\beta))\right)^{d/2}\right]$$

which is at most $\beta$ by the concentration bound (Theorem 6), as desired. So the claim holds: if $\rho$ is a good restriction then $f_\rho(x_T)$ is $\beta$-close to $\mathrm{sign}(p'(\rho))$. Thus to prove Lemma 3 it remains to show that at least a $1/(2C^d)$ fraction of all restrictions $\rho$ to $H$ are good.

We prove this in two steps. First we show (Lemma 4) that the polynomial $p'$ is not too concentrated: with probability at least $1/C^d$ over $\rho$, condition (i) of Definition 2 is satisfied. We then show (Lemma 5) that the polynomial $q$ is highly concentrated: the probability (over $\rho$) that condition (ii) is *not* satisfied is at most $1/(2C^d)$. Lemma 3 then follows by a union bound.

**Lemma 4.** *We have that $\mathbf{Pr}_\rho\left[|p'(\rho)| \geq t^*\right] \geq 1/C^d$.*

*Proof.* Using the fact that the critical index of $p$ is large, we will show that the polynomial $p'$ has large variance (close to 1), and hence we can apply the anti-concentration bound Theorem 7.

We start by establishing that $\mathbf{Var}[p']$ lies in the range $[1/2, 1]$. To see this, first recall that for $g : \{-1,1\}^n \to \mathbb{R}$ we have $\mathbf{Var}[g] = \sum_{\emptyset \neq S \subseteq [n]} \widehat{g}^2(S)$. It is thus clear that $\mathbf{Var}[p'] \leq \mathbf{Var}[p] = 1$. To establish the lower bound we use the property that the "tail" $T$ has "very small" influence in $p$, which is a consequence of the critical index of $p$ being large. More precisely, Lemma 1 yields

$$\sum_{i \in T} \mathrm{Inf}_i(p) \leq (1-\tau)^K \cdot \mathrm{Inf}(p) = (1-\tau)^{\alpha/\tau} \cdot \mathrm{Inf}(p) \leq d \cdot e^{-\alpha} \tag{2}$$

where the last inequality uses the fact that $\mathrm{Inf}(p) \leq d$. Therefore, we have:

$$\mathbf{Var}[p'] = \mathbf{Var}[p] - \sum_{T \cap S \neq \emptyset, S \subseteq [n]} \widehat{p}(S)^2 \geq 1 - \sum_{i \in T} \mathrm{Inf}_i(p) \geq 1 - de^{-\alpha} \geq 1/2$$

where the first inequality uses the fact that $\mathrm{Inf}_i(p) = \sum_{i \in S \subseteq [n]} \widehat{p}(S)^2$, the second follows from (2) and the third from our choice of $\alpha$. We have thus established that indeed $\mathbf{Var}[p'] \in [1/2, 1]$.

At this point, we would like to apply Theorem 7 for $p'$. Note however that $\mathbf{E}[p'] = \mathbf{E}[p] = \widehat{p}(\emptyset)$ which is not necessarily zero. To address this minor technical point we apply Theorem 7 twice: once for the polynomial $p'' = p' - \widehat{p}(\emptyset)$ and once for $-p''$. (Clearly, $\mathbf{E}[p''] = 0$ and $\mathbf{Var}[p''] = \mathbf{Var}[p'] \in [1/2, 1]$.) We thus get that, independent of the value of $\widehat{p}(\emptyset)$, we have $\mathbf{Pr}_\rho[|p'(\rho)| > 2^{-1/2} \cdot C^{-d}] \geq C^{-d}$, as desired. $\square$

**Lemma 5.** *We have that* $\mathbf{Pr}_\rho\big[\|q(\rho, x_T)\|_2 > t^* \cdot \big(\Theta(\log(1/\beta))\big)^{-d/2}\big] \leq 1/(2C^d)$.

*Proof.* To obtain the desired concentration bound we must show that the degree-$2d$ polynomial $Q(\rho) = \|q(\rho, x_T)\|_2^2$ has "small" variance. The desired bound then follows by an application of Theorem 6.

We thus begin by showing that $\|Q\|_2 \leq 3^d de^{-\alpha}$. To see this, we first note that $Q(\rho) = \sum_{\emptyset \neq S \subseteq T} \widehat{p}_\rho(S)^2$. Hence an application of the triangle inequality for norms and hypercontractivity (Theorem 5) yields:

$$\|Q\|_2 \leq \sum_{\emptyset \neq S \subseteq T} \|\widehat{p}_\rho(S)\|_4^2 \leq 3^d \sum_{\emptyset \neq S \subseteq T} \|\widehat{p}_\rho(S)\|_2^2.$$

We now proceed to bound from above the RHS term by term:

$$\sum_{\emptyset \neq S \subseteq T} \|\widehat{p}_\rho(S)\|_2^2 = \sum_{\emptyset \neq S \subseteq T} \mathbf{E}_\rho[\widehat{p}_\rho(S)^2] = \mathbf{E}_\rho\Big[\sum_{\emptyset \neq S \subseteq T} \widehat{p}_\rho(S)^2\Big] \leq \mathbf{E}_\rho\big[\mathrm{Inf}(p_\rho)\big] = \mathbf{E}_\rho\Big[\sum_{i \in T} \mathrm{Inf}_i(p_\rho)\Big]$$

$$= \sum_{i \in T} \mathbf{E}_\rho\big[\mathrm{Inf}_i(p_\rho)\big] = \sum_{i \in T} \mathrm{Inf}_i(p) \leq de^{-\alpha} \tag{3}$$

where the first inequality uses the fact $\mathrm{Inf}(p_\rho) \geq \sum_{\emptyset \neq S \subseteq T} \widehat{p}_\rho(S)^2$, the equality in (3) follows from Lemma 2, and the last inequality is Equation (2). We have thus shown that $\|Q\|_2 \leq 3^d de^{-\alpha}$.

We now upper bound $\mathbf{Pr}_\rho[Q(\rho) > (t^*)^2 \cdot \Theta(\log(1/\beta))^{-d}]$. Since $\|Q\|_2 \leq 3^d de^{-\alpha}$, Theorem 6 implies that for all $t > e^d$ we have $\mathbf{Pr}_\rho[Q(\rho) > t \cdot 3^d de^{-\alpha}] \leq \exp(-\Omega(t^{1/d}))$. Taking $t$ to be $\Theta(d^d \ln^d C)$ this upper bound is at most $1/(2C^d)$. Our choice of the parameter $\alpha$ gives $t \cdot d3^d \cdot e^{-\alpha} \leq (t^*)^2 \cdot \Theta(\log(1/\beta))^{-d}$. This completes the proof of Lemma 5, and thus also the proof of Lemma 3. $\square$

## 2.2 The small critical index case

In this section we show that if the critical index of $p$ is "small", then a random restriction of "few" variables causes $p$ to become regular with non-negligible probability. We do this by showing that no matter what the critical index is, a random restriction of all variables up to the $\tau$-critical index causes $p$ to become $\tau'$-regular, for some $\tau'$ not too much larger than $\tau$, with probability at least $1/(2C^d)$. More formally, we prove:

**Lemma 6.** *Let* $p : \{-1,1\}^n \to \mathbb{R}$ *be a degree-$d$ polynomial with $\tau$-critical index $k \in [n]$. Let $\rho$ be a random restriction that fixes $[k]$, and let $\tau' = (C' \cdot d \ln d \cdot \ln \frac{1}{\tau})^d \cdot \tau$ for some suitably large absolute constant $C'$. With probability at least $1/(2C^d)$ over the choice of $\rho$, the restricted polynomial $p_\rho$ is $\tau'$-regular.*

*Proof.* We must show that with probability at least $1/(2C^d)$ over $\rho$ the restricted polynomial $p_\rho$ satisfies

$$\text{Inf}_\ell(p_\rho) / \sum_{j=k+1}^n \text{Inf}_j(p_\rho) \leq \tau' \tag{4}$$

for all $\ell \in [k+1, n]$. Note that before the restriction, we have $\text{Inf}_\ell(p) \leq \tau \cdot \sum_{j=k+1}^n \text{Inf}_j(p)$ for all $\ell \in [k+1, n]$ because the $\tau$-critical index of $p$ is $k$.

Let us give an intuitive explanation of the proof. We first show (Lemma 7) that with probability at least $C^{-d}$ the denominator in (4) does not decrease under a random restriction. This is an anti-concentration statement that follows easily from Theorem 7. We then show (Lemma 8) that with probability at least $1 - C^{-d}/2$ the numerator in (4) does not increase by much under a random restriction, i.e. *no variable influence* $\text{Inf}_\ell(p_\rho), \ell \in [k+1, n]$, becomes too large. Thus both events occur (and $p_\rho$ is $\tau'$-regular) with probability at least $C^{-d}/2$.

We note that while each individual influence $\text{Inf}_\ell(p_\rho)$ is indeed concentrated around its expectation (see Claim 9), we need a concentration statement for $n - k$ such influences. This might seem difficult to achieve since we require bounds that are independent of $n$. We get around this difficulty by a "bucketing" argument that exploits the fact (at many different scales) that all but a few influences $\text{Inf}_\ell(p)$ must be "very small."

It remains to state and prove Lemmas 7 and 8. Consider the event $\mathcal{E} \stackrel{\text{def}}{=} \{\rho \in \{-1,1\}^k \mid \sum_{\ell=k+1}^n \text{Inf}_\ell(p_\rho) \geq \sum_{\ell=k+1}^n \text{Inf}_\ell(p)\}$. We first show:

**Lemma 7.** $\mathbf{Pr}_\rho[\mathcal{E}] \geq C^{-d}$.

*Proof.* It follows from Fact 14 and the Fourier expression of $\text{Inf}_\ell(p_\rho)$ that $A(\rho) \stackrel{\text{def}}{=} \sum_{\ell=k+1}^n \text{Inf}_\ell(p_\rho)$ is a degree-$2d$ polynomial. By Lemma 2 we get that $\mathbf{E}_\rho[A] = \sum_{\ell=k+1}^n \text{Inf}_\ell(p) > 0$. Also observe that $A(\rho) \geq 0$ for all $\rho \in \{-1,1\}^k$. We may now apply Theorem 7 to the polynomial $A' = A - \mathbf{E}_\rho[A]$, to obtain:

$$\mathbf{Pr}_\rho[\mathcal{E}] = \mathbf{Pr}[A' \geq 0] \geq \mathbf{Pr}[A' \geq C_0^{-2d} \cdot \sigma(A')] > C_0^{-2d} = C^{-d}. \qquad \square$$

We now turn to Lemma 8. Consider the event $\mathcal{J} \stackrel{\text{def}}{=} \{\rho \in \{-1,1\}^k \mid \max_{\ell \in [k+1,n]} \text{Inf}_\ell(p_\rho) > \tau' \sum_{j=k+1}^n \text{Inf}_j(p)\}$. We show:

**Lemma 8.** $\mathbf{Pr}_\rho[\mathcal{J}] \leq (1/2) \cdot C^{-d}$.

The rest of this subsection consists of the proof of Lemma 8. A useful intermediate claim is that the influences of individual variables do not increase by a lot under a random restriction (note that this claim, proved in Appendix B, does not depend on the value of the critical index):

**Claim 9.** *Let $p : \{-1,1\}^n \to \mathbb{R}$ be a degree-$d$ polynomial. Let $\rho$ be a random restriction fixing $[j]$. Fix any $t > e^{2d}$ and any $\ell \in [j+1, n]$. With probability at least $1 - \exp(-\Omega(t^{1/d}))$ over $\rho$, we have $\text{Inf}_\ell(p_\rho) \leq 3^d t \text{Inf}_\ell(p)$.*

Claim 9 says that for any given coordinate, the probability that its influence after a random restriction increases by a $t$ factor decreases exponentially in $t$. Note that Claim 9 and a naive union bound over all coordinates in $[k+1, n]$ does not suffice to prove Lemma 8. Instead, we proceed as follows: We partition the coordinates in $[k+1, n]$ into "buckets" according to their influence in the tail of $p$. In particular, the $i$-th bucket ($i \geq 0$) contains all variables $\ell \in [k+1, n]$ such that

$$\text{Inf}_\ell(p) / \sum_{j=k+1}^n \text{Inf}_j(p) \in [\tau/2^{i+1}, \tau/2^i].$$

We analyze the effect of a random restriction $\rho$ on the variables of each bucket $i$ separately and then conclude by a union bound over all the buckets.

So fix a bucket $i$. Note that, by definition, the number of variables in the $i$-th bucket is at most $2^{i+1}/\tau$. We bound from above the probability of the event $\mathcal{B}(i)$ that there exists a variable $\ell$ in bucket $i$ that violates the regularity constraint, i.e. such that $\mathrm{Inf}_\ell(p_\rho) > \tau' \sum_{\ell=k+1}^n \mathrm{Inf}_\ell(p)$. We will do this by a combination of Claim 9 and a union bound over the variables in the bucket. We will show:

**Claim 10.** *We have that* $\mathbf{Pr}_\rho[\mathcal{B}(i)] \leq 2^{-(i+2)} \cdot C^{-d}$.

The above claim completes the proof of Lemma 8 by a union bound across buckets. Indeed, assuming the claim, the probability that *any* variable $\ell \in [k+1, n]$ violates the condition $\mathrm{Inf}_\ell(p_\rho) \leq \tau' \sum_{\ell=k+1}^n \mathrm{Inf}_\ell(p)$ is at most $\sum_{i=0}^\infty \mathbf{Pr}_\rho[\mathcal{B}(i)] \leq C^{-d} 2^{-2} \sum_{i=0}^\infty (1/2)^i = (1/2) \cdot C^{-d}$.

It thus remains to prove Claim 10. Fix a variable $\ell$ in the $i$-th bucket. We apply Claim 9 selecting a value of $t = \widetilde{t} \stackrel{\text{def}}{=} (\ln \frac{C^d 4^{i+2}}{\tau})^d$. It is clear that $\widetilde{t} \leq c'^d (d + i + \ln \frac{1}{\tau})^d$ for some absolute constant $c'$. As a consequence, there is an absolute constant $C'$ such that for every $i$,

$$\widetilde{t} \leq 3^{-d} C'^d 2^i (d \ln d \ln \frac{1}{\tau})^d. \tag{5}$$

(To see this, note that for $i \leq 10 d \ln d$ we have $d + i + \ln \frac{1}{\tau} < 11 d \ln d \ln \frac{1}{\tau}$, from which the claimed bound is easily seen to hold. For $i > 10 d \ln d$, we use $d + i + \ln \frac{1}{\tau} < di \ln \frac{1}{\tau}$ and the fact that $i^d < 2^i$ for $i > 10 d \ln d$.)

Inequality (5) can be rewritten as $3^d \cdot \widetilde{t} \cdot \frac{\tau}{2^i} \leq \tau'$. Hence, our assumption on the range of $\mathrm{Inf}_\ell(p)$ gives

$$3^d \cdot \widetilde{t} \cdot \mathrm{Inf}_\ell(p) \leq \tau' \cdot \sum_{j=k+1}^n \mathrm{Inf}_j(p).$$

Therefore, by Claim 9, the probability that coordinate $\ell$ violates the condition $\mathrm{Inf}_\ell(p_\rho) \leq \tau' \sum_{j=k+1}^n \mathrm{Inf}_j(p)$ is at most $\tau/(C^d 4^{i+2})$ by our choice of $\widetilde{t}$. Since bucket $i$ contains at most $2^{i+1}/\tau$ coordinates, Claim 10 follows by a union bound. Hence Lemma 8, and thus Lemma 6, is proved. □

## 2.3 Putting Everything Together: Proof of Theorem 1

The following lemma combines the results of the previous two subsections (see Appendix B for a proof):

**Lemma 11.** *Let* $p : \{-1, 1\}^n \to \mathbb{R}$ *be a degree-$d$ polynomial and* $0 < \widetilde{\tau}, \beta < 1/2$. *Fix* $\alpha = \Theta(d \log \log(1/\beta) + d \log d)$ *and* $\widetilde{\tau}' = \widetilde{\tau} \cdot (C' d \ln d \ln(1/\widetilde{\tau}))^d$, *where $C'$ is a universal constant. (We assume w.l.o.g. that the variables are ordered s.t.* $\mathrm{Inf}_i(p) \geq \mathrm{Inf}_{i+1}(p)$, $i \in [n-1]$.) *One of the following statements holds true:*

1. *The polynomial $p$ is $\widetilde{\tau}$-regular.*

2. *With probability at least $1/(2C^d)$ over a random restriction $\rho$ fixing the first $\alpha/\widetilde{\tau}$ (most influential) variables of $p$, the function $\mathrm{sign}(p_\rho)$ is $\beta$-close to a constant function.*

3. *There exists a value $k \leq \alpha/\widetilde{\tau}$, such that with probability at least $1/(2C^d)$ over a random restriction $\rho$ fixing the first $k$ (most influential) variables of $p$, the polynomial $p_\rho$ is $\widetilde{\tau}'$-regular.*

*Proof of Theorem 1.* We begin by observing that any function $f$ on $\{-1, 1\}^n$ is equivalent to a decision tree where each internal node of the tree is labeled by a variable, every root-to-leaf path corresponds to a restriction $\rho$ that fixes the variables as they are set on the path, and every leaf is labeled with the restricted subfunction $f_\rho$. Given an arbitrary degree-$d$ PTF $f = \mathrm{sign}(p)$, we will construct a decision tree $\mathcal{T}$ of the form described in Theorem 1. It is clear that in any such tree every leaf function $f_\rho$ will be a degree-$d$ PTF; we must show that $\mathcal{T}$ has depth $\mathrm{depth}(d, \tau)$ and that with probability $1 - \tau$ over the choice of a random root-to-leaf path $\rho$, the restricted subfunction $f_\rho = \mathrm{sign}(p_\rho)$ is $\tau$-close to a $\tau$-regular degree-$d$ PTF.

For a tree $T$ computing $f = \mathrm{sign}(p)$, we denote by $N(T)$ its set of internal nodes and by $L(T)$ its set of leaves. We call a leaf $\rho \in L(T)$ "good" if the corresponding function $f_\rho$ is $\tau$-close to being $\tau$-regular. We call a leaf "bad" otherwise. Let $GL(T)$ and $BL(T)$ be the sets of "good" and "bad" leaves in $T$ respectively.

The basic approach for the proof is to invoke Lemma 11 repeatedly in a sequence of at most $2C^d \ln(1/\tau)$ stages. In the first stage we apply Lemma 11 to $f$ itself; this gives us an initial decision tree. In the second stage we apply Lemma 11 to those restricted subfunctions $f_\rho$ (corresponding to leaves of the initial decision tree) that are still $\tau$-far from being $\tau$-regular; this "grows" our initial decision tree. Subsequent stages continue similarly; we will argue that after at most $2C^d \ln(1/\tau)$ stages, the resulting tree satisfies the required properties for $\mathcal{T}$. In every application of Lemma 11 the parameters $\beta$ and $\widetilde{\tau}'$ are both taken to be $\tau$; note that taking $\widetilde{\tau}'$ to be $\tau$ sets the value of $\widetilde{\tau}$ in Lemma 11 to a value that is less than $\tau$.

We now provide the details. In the first stage, the initial application of Lemma 11 results in a tree $T_1$. This tree $T_1$ may consist of a single leaf node that is $\widetilde{\tau}$-regular (if $f$ is $\widetilde{\tau}$-regular to begin with – in this case, since $\widetilde{\tau} < \tau$, we are done), or a complete decision tree of depth $\alpha/\widetilde{\tau}$ (if $f$ had large critical index), or a complete decision tree of depth $k < \alpha/\widetilde{\tau}$ (if $f$ had small critical index). Note that in each case the depth of $T_1$ is at most $\alpha/\widetilde{\tau}$. Lemma 11 guarantees that:

$$\mathbf{Pr}_{\rho \in T_1}[\rho \in BL(T_1)] \leq 1 - 1/(2C^d),$$

where the probability is over a random root-to-leaf path $\rho$ in $T_1$.

In the second stage, the "good" leaves $\rho \in GL(T_1)$ are left untouched; they will be leaves in the final tree $\mathcal{T}$. For each "bad" leaf $\rho \in BL(T_1)$, we order the unrestricted variables in decreasing order of their influence in the polynomial $p_\rho$, and we apply Lemma 11 to $f_\rho$. This "grows" $T_1$ at each bad leaf by replacing each such leaf with a new decision tree; we call the resulting overall decision tree $T_2$.

A key observation is that the probability that a random path from the root reaches a "bad" leaf is significantly smaller in $T_2$ than in $T_1$; in particular

$$\mathbf{Pr}_{\rho \in T_2}[\rho \in BL(T_2)] \leq (1 - 1/(2C^d))^2.$$

We argue this as follows: Let $\rho$ be any fixed "bad" leaf in $T_1$, i.e. $\rho \in BL(T_1)$. The function $f_\rho$ is not $\widetilde{\tau}'$-regular and consequently not $\widetilde{\tau}$-regular. Thus, either statement (2) or (3) of Lemma 11 must hold when the Lemma is applied to $f_\rho$. The tree that replaces $\rho$ in $T_0$ has depth at most $\alpha/\widetilde{\tau}$, and a random root-to-leaf path $\rho_1$ *in this tree* reaches a "bad" leaf with probability at most $1 - 1/(2C^d)$. So the overall probability that a random root-to-leaf path in $T_2$ reaches a "bad" leaf is at most $(1 - 1/(2C^d))^2$.

Continuing in this fashion, in the $i$-th stage we replace all the bad leaves of $T_{i-1}$ by decision trees according to Lemma 11 and we obtain the tree $T_i$. An inductive argument gives that

$$\mathbf{Pr}_{\rho \in T_i}[\rho \in BL(T_i)] \leq (1 - 1/(2C^d))^i, \quad \text{which is at most } \tau \text{ for} \quad i^* \stackrel{\text{def}}{=} 2C^d \ln(1/\tau).$$

The depth of the overall tree will be the maximum number of stages ($2C^d \ln(1/\tau)$) times the maximum depth added in each stage (at most $\alpha/\widetilde{\tau}$, since we always restrict at most this many variables), which is at most $(\alpha/\widetilde{\tau}) \cdot i^*$. Since $\beta = \tau$, we get $\alpha = \Theta(d \log \log(1/\tau) + d \log d)$. Recalling that $\widetilde{\tau}'$ in Lemma 11 is set to $\tau$, we see that $\widetilde{\tau} = \tau/(C'd \ln d \ln(1/\tau))^{O(d)}$. By substitution we get that the depth of the tree is upper bounded by $d^{O(d)} \cdot (1/\tau) \cdot \log(1/\tau)^{O(d)}$ which concludes the proof of Theorem 1. $\square$

## 3 Every degree-$d$ PTF has a low-weight approximator

In this section we apply Theorem 1 to prove Theorem 2, which we restate below:

**Theorem 2** *Let $f(x) = \text{sign}(p(x))$ be any degree-$d$ PTF. Fix any $\epsilon > 0$ and let $\tau = (\Theta(1) \cdot \epsilon/d)^{8d}$. Then there is a polynomial $q(x)$ of degree $D = d + \text{depth}(d, \tau)$ and weight $n^d \cdot 2^{4\text{depth}(d,\tau)} \cdot (d/\epsilon)^{O(d)}$, which is such that the PTF $\text{sign}(q(x))$ is $\epsilon$-close to $f$.*

To prove Theorem 2, we first show that any sufficiently regular degree-$d$ PTF over $n$ variables has a low-weight approximator, of weight roughly $n^d$. Theorem 1 asserts that almost every leaf $\rho$ of $\mathcal{T}$ is close to a regular PTF; at each such leaf $\rho$ we use the low-weight approximator of the previous sentence to approximate the regular PTF, and thus to approximate $f_\rho$. Finally, we combine all of these low-weight polynomials to get an overall PTF of low weight which is a good approximator for $f$. We give details below.

### 3.1 Low-weight approximators for regular PTFs

In this subsection we prove that every sufficiently regular PTF has a low-weight approximator of degree $d$:

**Lemma 12.** *Given $\epsilon > 0$, let $\tau = (\Theta(1) \cdot \epsilon/d)^{8d}$. Let $p : \{-1,1\}^n \to \mathbb{R}$ be a $\tau$-regular degree-$d$ polynomial with $\mathbf{Var}[p] = 1$. There exists a degree-$d$ polynomial $q : \{-1,1\}^n \to \mathbb{R}$ of weight $n^d \cdot (d/\epsilon)^{O(d)}$ such that $\mathrm{sign}(q(x))$ is an $\epsilon$-approximator for $\mathrm{sign}(p(x))$.*

*Proof.* The polynomial $q$ is obtained by rounding the weights of $p$ to an appropriate granularity, similar to the regular case in [Ser07] for the $d=1$ case. To show that this works, we use the fact that regular PTFs have very good anti-concentration. In particular we will use the following claim, which is proved using the invariance principle [MOO05] and Gaussian anti-concentration [CW01] (see Appendix D for the proof):

**Claim 13.** *Let $p : \{-1,1\}^n \to \mathbb{R}$ be a $\tau$-regular degree-$d$ polynomial with $\mathbf{Var}[p] = 1$. Then $\mathbf{Pr}_x[|p(x)| \leq \tau] \leq O(d\tau^{1/8d})$.*

We turn to the detailed proof of Lemma 12. We first note that if the constant coefficient $\widehat{p}(\emptyset)$ of $P$ has magnitude greater than $(O(\log(1/\epsilon)))^{d/2}$, then Theorem 6 (applied to $p(x) - \widehat{p}(\emptyset)$) implies that $\mathrm{sign}(p(x))$ agrees with $\mathrm{sign}(\widehat{p}(\emptyset))$ for at least a $1 - \epsilon$ fraction of inputs $x$. So in this case $\mathrm{sign}(p(x))$ is $\epsilon$-close to a constant function, and the conclusion of the Lemma certainly holds. Thus we henceforth assume that $|\widehat{p}(\emptyset)|$ is at most $(O(\log(1/\epsilon)))^{d/2}$.

Let
$$\alpha = \tau / (Kn \cdot \ln(4/\epsilon))^{d/2}$$
where $K > 0$ is an absolute constant (specified later). For each $S \neq \emptyset$ let $\widehat{q}(S)$ be the value obtained by rounding $\widehat{p}(S)$ to the nearest integer multiple of $\alpha$, and let $\widehat{q}(\emptyset)$ equal $\widehat{p}(\emptyset)$. This defines a degree-$d$ polynomial $q(x) = \sum_S \widehat{q}(S) \chi_S(x)$. It is easy to see that rescaling by $\alpha$, all of the non-constant coefficients of $q(x)/\alpha$ are integers. Since each coefficient $\widehat{q}(S)$ has magnitude at most twice that of $\widehat{p}(S)$, we may bound the sum of squares of coefficients of $q(x)/\alpha$ by

$$\frac{\widehat{p}(\emptyset)^2}{\alpha^2} + \frac{\sum_{S \neq \emptyset} 4\widehat{p}(S)^2}{\alpha^2} \leq \frac{(O(\log(1/\epsilon)))^d}{\alpha^2} \leq n^d \cdot (d/\epsilon)^{O(d)}.$$

We now observe that the constant coefficient $\widehat{p}(\emptyset)$ of $q(x)$ can be rounded to an integer multiple of $\alpha$ without changing the value of $\mathrm{sign}(q(x))$ for any input $x$. Doing this, we obtain a polynomial $q'(x)/\alpha$ with all integer coefficients, weight $n^d \cdot (d/\epsilon)^{O(d)}$, and which has $\mathrm{sign}(q'(x)) = \mathrm{sign}(q(x))$ for all $x$.

In the rest of our analysis we shall consider the polynomial $q(x)$ (recall that the constant coefficient of $q(x)$ is precisely $\widehat{p}(\emptyset)$). It remains to show that $\mathrm{sign}(q)$ is an $\epsilon$-approximator for $\mathrm{sign}(p)$. For each $S \neq \emptyset$ let $\widehat{e}(S)$ equal $\widehat{p}(S) - \widehat{q}(S)$. This defines a polynomial (with constant term 0) $e(x) = \sum_S \widehat{e}(S) x_S$, and we have $q(x) + e(x) = p(x)$. (The coefficients $\widehat{e}(S)$ are the "errors" induced by approximating $\widehat{p}(S)$ by $\widehat{q}(S)$.)

Recall that $\tau = (\Theta(1) \cdot \epsilon/d)^{8d}$. For any input $x$, we have that $\mathrm{sign}(q(x)) \neq \mathrm{sign}(p(x))$ only if either (i) $|e(x)| \geq \tau$, or (ii) $|p(x)| \leq \tau$. Since each coefficient of $e(x)$ satisfies $|\widehat{e}(S)| \leq \alpha/2 \leq \frac{\tau}{2(Kn \cdot \ln(4/\epsilon))^{d/2}}$, the sum of squares of all (at most $n^d$) coefficients of $e$ is at most

$$\sum_S \widehat{e}(S)^2 \leq \frac{\tau^2}{4(K \ln(4/\epsilon))^d}, \quad \text{and thus} \quad \|e\| \leq \frac{\tau}{2(K \ln(4/\epsilon))^{d/2}}.$$

Applying Theorem 6, we get that $\mathbf{Pr}_x[|e(x)| \geq \tau] \leq \epsilon/2$ (for a suitable absolute constant choice of $K$), so we have upper bounded the probability of (i).

For (ii), we use the anti-concentration bound for regular polynomials, Claim 13. This directly gives us that $\mathbf{Pr}_x[|p(x)| \leq \tau] \leq O(d\tau^{1/8d}) \leq \epsilon/2$.

Thus the probability, over a random $x$, that either (1) or (2) holds is at most $\epsilon$. Consequently $\text{sign}(q)$ is an $\epsilon$-approximator for $\text{sign}(p)$, and Lemma 12 is proved. □

## 3.2 Proof of Theorem 2

Let $f = \text{sign}(p)$ be an arbitrary degree-$d$ PTF over $n$ Boolean variables, and let $\epsilon > 0$ be the desired approximation parameter. We invoke Theorem 1 with its "$\tau$" parameter set to $\tau = (\Theta(1) \cdot (\epsilon/2)/d)^{8d}$ (i.e. our choice of $\tau$ is obtained by plugging in "$\epsilon/2$" for $\epsilon$ in the first sentence of Lemma 12). For each leaf $\rho$ of the tree $T$ as described in Theorem 1 (we call these "good" leaves), let $g^{(\rho)}$ be a $\tau$-regular degree-$d$ PTF that is $\tau$-close to $f_\rho$. By Lemma 12, for each such leaf $\rho$ there is a degree-$d$ polynomial of weight $n^d \cdot (d/\epsilon)^{O(d)}$, which we denote $q^{(\rho)}$, such that $g^{(\rho)}$ is $\epsilon/2$-close to $\text{sign}(q^{(\rho)})$. For each of the other leaves in $T$ (which are reached by at most a $\tau$ fraction of all inputs to $T$ – we call these "bad" leaves), for which $f_\rho$ is not $\tau$-close to any $\tau$-regular degree-$d$ PTF, let $q^{(\rho)}$ be the constant-1 function.

For each leaf $\rho$ of depth $r$ in $T$, let $P_\rho(x)$ be the unique multilinear polynomial of degree $r$ which outputs $2^r$ iff $x$ reaches $\rho$ and outputs 0 otherwise. (As an example, if $\rho$ is a leaf which is reached by the path "$x_3 = -1, x_6 = 1, x_2 = 1$" from the root in $T$, then $P_\rho(x)$ would be $(1 - x_3)(1 + x_6)(1 + x_2)$.) Our final PTF is

$$g(x) = \text{sign}(Q(x)), \quad \text{where} \quad Q(x) = \sum_\rho P_\rho(x) q^{(\rho)}(x).$$

It is easy to see that on any input $x$, the value $Q(x)$ equals $2^{|\rho_x|} \cdot q^{(\rho_x)}(x)$, where we write $\rho_x$ to denote the leaf of $T$ that $x$ reaches and $|\rho_x|$ to denote the depth of that leaf. Thus $\text{sign}(Q(x))$ equals $\text{sign}(q^{(\rho_x)}(x))$ for each $x$, and from this it follows that $\mathbf{Pr}_x[g(x) \neq f(x)]$ is at most $\tau + \tau + \epsilon/2 < \epsilon$. Here the first $\tau$ is because a random input $x$ may reach a bad leaf with probability up to $\tau$; the second $\tau$ is because for each good leaf $\rho$, the function $g^{(\rho)}$ is $\tau$-close to $f_\rho$; and the $\epsilon/2$ is because $\text{sign}(q^{(\rho)})$ is $\epsilon/2$-close to $g^{(\rho)}$.

Since $T$ has depth $\text{depth}(d, \tau)$, it is easy to see that $Q$ has degree at most $\text{depth}(d, \tau) + d$. It is clear that the coefficients of $Q$ are all integers, so it remains only to bound the sum of squares of these coefficients. Each polynomial addend $P_\rho(x) q^{(\rho)}(x)$ in the sum is easily seen to have sum of squared coefficients

$$\sum_S \widehat{P_\rho q^{(\rho)}}(S)^2 = \mathbf{E}[(P_\rho \cdot q^{(\rho)})^2] \leq \left(\max_x P_\rho(x)^2\right) \cdot \mathbf{E}[q^{(\rho)}(x)^2] \leq 2^{2\text{depth}(d,\tau)} \cdot n^d \cdot (d/\epsilon)^{O(d)}. \quad (6)$$

Since $T$ has depth $\text{depth}(d, \tau)$, the number of leaves $\rho$ is at most $2^{\text{depth}(d,\tau)}$, and hence for each $S$ by Cauchy-Schwarz we have

$$\widehat{Q}(S)^2 = \left(\sum_\rho \widehat{P_\rho q^{(\rho)}}(S)\right)^2 \leq 2^{\text{depth}(d,\tau)} \cdot \sum_\rho \widehat{P_\rho q^{(\rho)}}(S)^2. \quad (7)$$

This implies that the total weight of $Q$ is

$$\begin{aligned} \sum_S \widehat{Q}(S)^2 &\leq 2^{\text{depth}(d,\tau)} \cdot \sum_{\rho,S} \widehat{P_\rho q^{(\rho)}}(S)^2 & \text{(using (7))} \\ &\leq 2^{2\text{depth}(d,\tau)} \max_\rho \left(\sum_S \widehat{P_\rho q^{(\rho)}}(S)^2\right) \\ &\leq 2^{4\text{depth}(d,\tau)} \cdot n^d \cdot (d/\epsilon)^{O(d)}, & \text{(using (6))} \end{aligned}$$

and Theorem 2 is proved. □

## 3.3 Degree-$d$ PTFs require $\widetilde{\Omega}(n^d)$-weight approximators

In this section we give two lower bounds on the weight required to $\epsilon$-approximate certain degree-$d$ PTFs. (We use the notation $\Omega_d()$ below to indicate that the hidden constant of the big-Omega depends on $d$.)

**Theorem 3.** *For all sufficiently large $n$, there is a degree-$d$ $n$-variable PTF $f(x)$ with the following property: Let $K(d)$ be any positive-valued function depending only on $d$. Suppose that $g(x) = \mathrm{sign}(q(x))$ is a degree-$K(d)$ PTF with integer coefficients $\widehat{q}(S)$ such that $\mathrm{dist}(f,g) \leq \epsilon^\star$ where $\epsilon^\star \stackrel{def}{=} C^{-d}/2$. Then the weight of $q$ is $\Omega_d(n^d/\log n)$.*

**Theorem 4.** *For all sufficiently large $n$, there is a degree-$d$ $n$-variable PTF $f(x)$ with the following property: Suppose that $g(x) = \mathrm{sign}(q(x))$ is any PTF (of any degree) with integer coefficients $\widehat{q}(S)$ such that $\mathrm{dist}(f,g) \leq \epsilon^\star$ where $\epsilon^\star \stackrel{def}{=} C^{-d}/2$. Then the weight of $q$ is $\Omega_d(n^{d-1})$.*

Viewing $d$ and $\epsilon$ as constants, Theorem 3 implies that the $O(n^d)$ weight bound of our $\epsilon$-approximator from Theorem 2 (which has constant degree) is essentially optimal for any constant-degree $\epsilon$-approximator. Theorem 4 says that there is only small room for improving our weight bound even if arbitrary-degree PTFs are allowed as approximators. We prove these results in Appendix D.2.

# A  Useful Background Results

## A.1  Fourier Analysis over $\{-1,1\}^n$.

We consider functions $f : \{-1,1\}^n \to \mathbb{R}$ (though we often focus on Boolean-valued functions which map to $\{-1,1\}$), and we think of the inputs $x$ to $f$ as being distributed according to the uniform probability distribution. The set of such functions forms a $2^n$-dimensional inner product space with inner product given by $\langle f,g \rangle = \mathbf{E}[f(x)g(x)]$. The set of functions $(\chi_S)_{S \subseteq [n]}$ defined by $\chi_S(x) = \prod_{i \in S} x_i$ forms a complete orthonormal basis for this space. Given a function $f : \{-1,1\}^n \to \mathbb{R}$ we define its *Fourier coefficients* by $\widehat{f}(S) \stackrel{\text{def}}{=} \mathbf{E}[f(x)\chi_S(x)]$, and we have that $f(x) = \sum_S \widehat{f}(S)\chi_S(x)$. We refer to the maximum $|S|$ over all nonzero $\widehat{f}(S)$ as the *Fourier degree* of $f$.

As an easy consequence of orthonormality we have *Plancherel's identity* $\langle f,g \rangle = \sum_S \widehat{f}(S)\widehat{g}(S)$, which has as a special case *Parseval's identity*, $\mathbf{E}[f(x)^2] = \sum_S \widehat{f}(S)^2$. From this it follows that for every $f : \{-1,1\}^n \to \{-1,1\}$ we have $\sum_S \widehat{f}(S)^2 = 1$. We recall the well-known fact (see e.g. [KKL88]) that the total influence $\mathrm{Inf}(f)$ of any Boolean function equals $\sum_S \widehat{f}(S)^2 |S|$. Note that, in this setting, the expectation and the variance can be expressed in terms of the Fourier coefficients of $f$ by $\mathbf{E}[f] = \widehat{f}(\emptyset)$ and $\mathbf{Var}[f] = \sum_{\emptyset \neq S \subseteq [n]} \widehat{f}(S)^2$.

## A.2  Useful Probability Bounds for Section 2

We first recall the following moment bound for low-degree polynomials, which is equivalent to the well-known hypercontractive inequality of [Bon70, Gro75]:

**Theorem 5.** *Let $p : \{-1,1\}^n \to \mathbb{R}$ be a degree-$d$ polynomial and $q > 2$. Then*

$$\|p\|_q \le (q-1)^{d/2} \|p\|_2.$$

The following concentration bound for low-degree polynomials, a simple corollary of hypercontractivity, is well known (see e.g. [O'D07b, DFKO06, AH09]):

**Theorem 6.** *Let $p : \{-1,1\}^n \to \mathbb{R}$ be a degree-$d$ polynomial. For any $t > e^d$, we have*

$$\mathbf{Pr}_x[|p(x)| \ge t\|p\|_2] \le \exp(-\Omega(t^{2/d})).$$

We will also need the following weak anti-concentration bound for low-degree polynomials over the cube:

**Theorem 7** ([DFKO06, AH09]). *There is a universal constant $C_0 > 1$ such that for any non-zero degree-d polynomial $p : \{-1,1\}^n \to \mathbb{R}$ with $\mathbf{E}[p] = 0$, we have*

$$\mathbf{Pr}_x[p(x) > C_0^{-d} \cdot \|p\|_2] > C_0^{-d}.$$

Throughout this paper, we let $C = C_0^2$, where $C_0$ is the universal constant from Theorem 7. Note that since $C > C_0$, Theorem 7 holds for $C$ as well.

## A.3 Useful Probability Bounds for Section 3

We denote by $\mathcal{N}^n$ the standard $n$-dimensional Gaussian distribution $\mathcal{N}(0,1)^n$.

The following two facts will be useful in the proof of Theorem 2, in particular in the analysis of the regular case. The first fact is a powerful anti-concentration bound for low-degree polynomials over independent Gaussian random variables:

**Theorem 8** ([CW01]). *Let $p : \mathbb{R}^n \to \mathbb{R}$ be a nonzero degree-d polynomial. For all $\epsilon > 0$ we have*

$$\mathbf{Pr}_{\mathcal{G} \sim \mathcal{N}^n}[|p(\mathcal{G})| \leq \epsilon \|p\|_2] \leq O(d\epsilon^{1/d}).$$

We note that the above bound is essentially tight, even for multi-linear polynomials.

The second fact is a version of the invariance principle of Mossel, O'Donnell and Oleszkiewicz, specifically Theorem 3.19 under hypothesis **H4** in [MOO05]:

**Theorem 9** ([MOO05]). *Let $p(x) = \sum_{S \subseteq [n], |S| \leq d} \widehat{p}(S) \chi_S(x)$ be a degree-d multilinear polynomial with $\mathbf{Var}[p] = 1$. Suppose each coordinate $i \in [n]$ has $\mathrm{Inf}_i(p) \leq \tau$. Then,*

$$\sup_{t \in \mathbb{R}} |\mathbf{Pr}_x[p(x) \leq t] - \mathbf{Pr}_{\mathcal{G} \sim \mathcal{N}^n}[p(\mathcal{G}) \leq t]| \leq O(d\tau^{1/(8d)}).$$

# B Omitted Proofs from Section 2

## B.1 Proof of Lemma 1

Recall Lemma 1:

**Lemma 1.** *Let $p : \{-1,1\}^n \to \mathbb{R}$ and $\tau > 0$. Let $k$ be the $\tau$-critical index of $p$. For $j \in [0,k]$ we have*

$$\sum_{i=j+1}^n \mathrm{Inf}_i(p) \leq (1-\tau)^j \cdot \mathrm{Inf}(p).$$

*Proof.* The lemma trivially holds for $j = 0$. In general, since $j$ is at most $k$, we have that $\mathrm{Inf}_j(p) \geq \tau \cdot \sum_{i=j}^n \mathrm{Inf}_i(p)$, or equivalently $\sum_{i=j+1}^n \mathrm{Inf}_i(p) \leq (1-\tau) \cdot \sum_{i=j}^n \mathrm{Inf}_i(p)$ which yields the claimed bound. □

## B.2 Proof of Lemma 2

To prove Lemma 2, we first recall an observation about the expected value of Fourier coefficients under random restrictions (see e.g. [LMN93]):

**Fact 14.** *Let $p : \{-1,1\}^n \to \mathbb{R}$. Consider a random assignment $\rho$ to the variables $x_1, \ldots, x_k$. Fix any $S \subseteq [k+1, n]$. Then we have $\widehat{p_\rho}(S) = \sum_{T \subseteq [k]} \widehat{p}(S \cup T) \rho_T$ and therefore $\mathbf{E}_\rho[\widehat{p_\rho}(S)^2] = \sum_{T \subseteq [k]} \widehat{p}(S \cup T)^2$.*

In words, the above fact says that all the Fourier weight on sets of the form $S \cup \{$any subset of restricted variables$\}$ "collapses" down onto $S$ in expectation. Consequently, the influence of an unrestricted variable does not change in expectation under random restrictions:

**Lemma 2.** *Let $p : \{-1,1\}^n \to \mathbb{R}$. Consider a random assignment $\rho$ to the variables $x_1, \ldots, x_k$ and fix $\ell \in [k+1, n]$. Then $\mathbf{E}_\rho[\mathrm{Inf}_\ell(p_\rho)] = \mathrm{Inf}_\ell(p)$.*

*Proof.* We have

$$\mathbf{E}_\rho[\mathrm{Inf}_\ell(p_\rho)] = \mathbf{E}_\rho\Big[\sum_{\ell \in S \subseteq [k+1,n]} \widehat{p_\rho}(S)^2\Big] = \sum_{T \subseteq [k]} \sum_{\ell \in S \subseteq [k+1,n]} \widehat{p}(S \cup T)^2$$
$$= \sum_{\ell \in U \subseteq [n]} \widehat{p}(U)^2 = \mathrm{Inf}_\ell(p).$$

□

## B.3 Proof of Claim 9

Recall Claim 9:

**Claim 9.** *Let $p : \{-1,1\}^n \to \mathbb{R}$ be a degree-$d$ polynomial. Let $\rho$ be a random restriction fixing $[j]$. Fix any $t > e^{2d}$ and any $\ell \in [j+1, n]$. With probability at least $1 - \exp(-\Omega(t^{1/d}))$ over $\rho$, we have $\mathrm{Inf}_\ell(p_\rho) \le 3^d t \mathrm{Inf}_\ell(p)$.*

*Proof.* The identity $\mathrm{Inf}_\ell(p_\rho) = \sum_{\ell \in S \subseteq [j+1,n]} \widehat{p_\rho}(S)^2$ and Fact 14 imply that $\mathrm{Inf}_\ell(p_\rho)$ is a degree-$2d$ polynomial in $\rho$. Hence the claim follows from the concentration bound, Theorem 6, assuming we can appropriately upper bound the $l_2$ norm of the polynomial $\mathrm{Inf}_\ell(p_\rho)$. So to prove Claim 9 it suffices to show that

$$\|\mathrm{Inf}_\ell(p_\rho)\|_2 \le 3^d \mathrm{Inf}_\ell(p). \tag{8}$$

The proof of Equation (8) is similar to the argument establishing that $\|Q\|_2 \le 3^d d e^{-\alpha}$ in Section 2.1. The triangle inequality tells us that we may bound the $l_2$-norm of each squared-coefficient separately:

$$\|\mathrm{Inf}_\ell(p_\rho)\|_2 \le \sum_{\ell \in S \subseteq [j+1,n]} \|\widehat{p_\rho}(S)^2\|_2.$$

Since $\widehat{p_\rho}(S)$ is a degree-$d$ polynomial, Theorem 5 yields that

$$\|\widehat{p_\rho}(S)^2\|_2 = \|\widehat{p_\rho}(S)\|_4^2 \le 3^d \|\widehat{p_\rho}(S)\|_2^2,$$

hence

$$\|\mathrm{Inf}_\ell(p_\rho)\|_2 \le 3^d \sum_{\ell \in S \subseteq [j+1,n]} \|\widehat{p_\rho}(S)\|_2^2 = 3^d \mathrm{Inf}_\ell(p),$$

where the last equality is a consequence of Fact 14. Thus Equation (8) holds, and Claim 9 is proved. □

## B.4 Proof of Lemma 11

Recall Lemma 11:

**Lemma 11.** *Let $p : \{-1, 1\}^n \to \mathbb{R}$ be a degree-$d$ polynomial and $0 < \widetilde{\tau}, \beta < 1/2$. Fix $\alpha = \Theta(d \log \log(1/\beta) + d \log d)$ and $\widetilde{\tau}' = \widetilde{\tau} \cdot (C' d \ln d \ln(1/\widetilde{\tau}))^d$, where $C'$ is a universal constant. (We assume w.l.o.g. that the variables are ordered s.t. $\mathrm{Inf}_i(p) \geq \mathrm{Inf}_{i+1}(p)$, $i \in [n-1]$.) One of the following statements holds true:*

1. *The polynomial $p$ is $\widetilde{\tau}$-regular.*

2. *With probability at least $1/(2C^d)$ over a random restriction $\rho$ fixing the first $\alpha/\widetilde{\tau}$ (most influential) variables of $p$, the function $\mathrm{sign}(p_\rho)$ is $\beta$-close to a constant function.*

3. *There exists a value $k \leq \alpha/\widetilde{\tau}$, such that with probability at least $1/(2C^d)$ over a random restriction $\rho$ fixing the first $k$ (most influential) variables of $p$, the polynomial $p_\rho$ is $\widetilde{\tau}'$-regular.*

*Proof.* The proof is by a case analysis based on the value $\ell$ of the $\widetilde{\tau}$-critical index of the polynomial $p$. If $\ell = 0$, then by definition $p$ is $\widetilde{\tau}$-regular, hence the first statement of the lemma holds. If $\ell > \alpha/\widetilde{\tau}$, then we randomly restrict the first $\alpha/\widetilde{\tau}$ many variables. Lemma 3 says that for a random restriction $\rho$ fixing these variables, with probability at least $1/(2C^d)$ the (restricted) degree-$d$ PTF $\mathrm{sign}(p_\rho)$ is $\beta$-close to a constant. Hence, in this case, the second statement holds. To handle the case $\ell \in [1, \alpha/\widetilde{\tau}]$, we apply Lemma 6. This lemma says that with probability at least $1/(2C^d)$ over a random restriction $\rho$ fixing variables $[\ell]$, the polynomial $p_\rho$ is $\widetilde{\tau}'$-regular, so the third statement of Lemma 11 holds. □

## C Comparison with [MZ09]

We comment on the relation of our main result, Theorem 1, with a very similar decomposition result of Meka and Zuckerman [MZ09]. They obtain a small-depth decision tree such that most leaves are $\epsilon$-close to being $\epsilon$-regular under a *stronger* definition of regularity, which we will call "$\epsilon$-regularity in $l_2$" to distinguish it from our notion.

Let $p : \{-1, 1\}^n \to \mathbb{R}$ be a polynomial and $\epsilon > 0$. We say that the polynomial $p$ is "$\epsilon$-regular in $l_2$" if

$$\sqrt{\sum_{i=1}^{n} \mathrm{Inf}_i(p)^2} \leq \epsilon \cdot \sum_{i=1}^{n} \mathrm{Inf}_i(p).$$

Recall that in our definition of regularity, instead of upper bounding the $l_2$-norm of the influence vector $I = (\mathrm{Inf}_1(p), \ldots, \mathrm{Inf}_n(p))$ by $\epsilon$ times the total influence of $p$ (i.e. the $l_1$ norm of $I$), we upper bound the $l_\infty$ norm (i.e. the maximum influence). We may thus call our notion "$\epsilon$-regularity in $l_\infty$".

Note that if a polynomial is $\epsilon$-regular in $l_2$, then it is also $\epsilon$-regular in $l_\infty$. (And this implication is easily seen to be essentially tight, e.g. if we have many variables with tiny influence and one variable with an $\epsilon$-fraction of the total influence.) For the other direction, if a polynomial is $\epsilon$-regular in $l_\infty$, then it is $\sqrt{\epsilon}$-regular in $l_2$. (This is also tight if we have $1/\epsilon$ many variables with influence $\epsilon$.)

Meka and Zuckerman prove the following statement:

> Every degree-$d$ PTF $f = \mathrm{sign}(p)$ can be expressed as a decision tree of depth $2^{O(d)} \cdot (1/\epsilon^2) \log^2(1/\epsilon)$ with variables at the internal nodes and a degree-$d$ PTF $f_\rho = \mathrm{sign}(p_\rho)$ at each leaf $\rho$, such that with probability $1 - \epsilon$, a random root-to-leaf path reaches a leaf $\rho$ such that $f_\rho$ is $\epsilon$-close to being $\epsilon$-regular in $l_2$. (In particular, for a "good" leaf $\rho$, either $p_\rho$ will be $\epsilon$-regular in $l_2$ or $f_\rho$ will be $\epsilon$-close to a constant).

Theorem 1 shows exactly the same statement as the one above if we replace "$l_2$" by "$l_\infty$" and the depth of the tree by $(1/\epsilon) \cdot (d\log(1/\epsilon))^{O(d)}$.

Since $\epsilon$-regularity in $l_2$ implies $\epsilon$-regularity in $l_\infty$, the result of [MZ09] implies a version of Theorem 1 which has depth of $2^{O(d)} \cdot (1/\epsilon^2) \log^2(1/\epsilon)$. Hence the [MZ09] result and our result are quantitatively incomparable to each other. Roughly, if $d$ is a constant (independent of $\epsilon$), then our Theorem 1 is asymptotically better when $\epsilon$ becomes small. This range of parameters is quite natural in the context of pseudo-random generators. In particular, in the recent proof that $\text{poly}(1/\epsilon)$-wise independence $\epsilon$-fools degree-2 PTFs [DKN09], using [MZ09] instead of Theorem 1, would give a worse bound on the degree of independence (namely, $\widetilde{O}(\epsilon^{-18})$ as opposed to $\widetilde{O}(\epsilon^{-9})$). On the other hand, if $d = \widetilde{\Omega}(\log(1/\epsilon))$, then the result of [MZ09] is better.

## D   Omitted Proofs from Section 3

### D.1   Proof of Claim 13

**Claim 13.** *Let $p : \{-1,1\}^n \to \mathbb{R}$ be a $\tau$-regular degree-$d$ polynomial with $\mathbf{Var}[p] = 1$. Then $\mathbf{Pr}_x[|p(x)| \le \tau] \le O(d\tau^{1/8d})$.*

*Proof.* We recall that, since $\mathbf{Var}[p] = 1$ and $p$ is of degree $d$, it holds $\text{Inf}(p) \le d$. Thus, since $p$ is $\tau$-regular, we have that $\max_{i \in [n]} \text{Inf}_i(p) \le d\tau$. An application of the invariance principle (Theorem 9) in tandem with anti-concentration in gaussian space (Theorem 8) yields

$$\begin{aligned}\mathbf{Pr}_x[|p(x)| \le \tau] &\le O(d \cdot (d\tau)^{1/8d}) + \mathbf{Pr}_{\mathcal{G} \sim \mathcal{N}^n}[|p(\mathcal{G})| \le \tau] \\ &\le O(d\tau^{1/8d}) + O(d\tau^{1/d}) = O(d\tau^{1/8d}),\end{aligned}$$

and the claim follows. $\square$

### D.2   Proof of Lower Bound Theorems 3 and 4

Theorems 3 and 4 are both consequences of the following theorem:

**Theorem 10.** *There exists a set $\mathcal{C} = \{f_1, \ldots, f_M\}$ of $M = 2^{\Omega_d(n^d)}$ degree-$d$ PTFs $f_i$ such that for any $1 \le i < j \le M$, we have $\text{dist}(f_i, f_j) \ge C^{-d}$.*

**Proof of Theorems 3 and 4 assuming Theorem 10:** First we prove Theorem 3. We begin by claiming that there are at most $\left(3\binom{n}{\le K(d)}\right)^A$ many integer-weight PTFs of degree $K(d)$ and weight at most $A$. This is because any such PTF can be obtained by making a sequence of $A$ steps, where at each step either $-1$, $0$, or $1$ is added to one of the $\binom{n}{\le K(d)}$ many monomials of degree at most $K(d)$. Each step can be carried out in $3\binom{n}{\le K(d)}$ ways, giving the claimed bound.

By Theorem 10, there are $M$ distinct degree-$d$ PTFs $f_1, \ldots, f_M$ any two of which are $C^{-d}$-far from each other. Consequently any Boolean function (in particular, any weight-$A$ degree-$K(d)$ PTF $g$) can have $\text{dist}(g, f_i) \le C^{-d}/2$ for at most one $f_i$. Since there are only $\left(3\binom{n}{\le K(d)}\right)^A$ many weight-$A$ degree-$K(d)$ PTFs, and $\left(3\binom{n}{\le K(d)}\right)^A$ is less than $M$ for some $A = \Omega_d(n^d/\log n)$, it follows that some $f_i$ must have distance at least $C^{-d}/2$ from every weight-$A$, degree-$K(d)$ PTF. This gives Theorem 3.

The proof of Theorem 4 is nearly identical. We now use the fact that there are at most $(3 \cdot 2^n)^A$ many integer-weight PTFs of weight at most $A$ (and any degree), and use the fact that $(3 \cdot 2^n)^A$ is less than $M$ for some $A = \Omega_d(n^{d-1})$. $\square$

It remains to prove Theorem 10.

### D.2.1 Proof of Theorem 10

The proof is by the probabilistic method. We define the following distribution $\mathcal{D}$ over $n$-variable degree-$d$ polynomials. A draw of $p(x) = \sum_{S \subset [n], |S|=d} \widehat{p}(S) \chi_S(x)$ from $\mathcal{D}$ is obtained in the following way: each of the $\binom{n}{d}$ coefficients $\widehat{p}(S)$ is independently and uniformly selected from $\{-1, 1\}$.

We will prove Theorem 10 using Lemma 15, which says that it is extremely likely the polynomial $c$ – the product of two independent draws $a$ and $b$ from $\mathcal{D}$ – will have both small bias and large variance.

**Lemma 15.** *Let $a(x)$ and $b(x)$ be two degree-$d$ polynomials drawn independently from $\mathcal{D}$, and let $c(x) = a(x)b(x)$. Then with probability at least $1 - 2^{-\Omega_d(n^d)}$ we have:*

1. $|\widehat{c}(\emptyset)| \leq \frac{1}{4} C^{-d} \binom{n/2}{d}$, and

2. $\mathbf{Var}[c] \stackrel{\text{def}}{=} \sum_{|S|>0} \widehat{c}(S)^2 \geq \frac{1}{12} \binom{n/2}{d}^2$.

Suppose that Lemma 15 holds. Let $a(x)$ and $b(x)$ be independent draws from $\mathcal{D}$ and let $c(x) = a(x)b(x)$ which satisfies the conclusions of the lemma. Then the constant term $\widehat{c}(\emptyset)$ is small compared with the variance of $c(x)$. Let us rescale so the variance is 1; i.e. define the polynomial

$$e(x) \stackrel{\text{def}}{=} \frac{c(x)}{\mathbf{Var}[c]^{1/2}}$$

so $\mathbf{Var}[e] = 1$ and $|\widehat{e}(\emptyset)| < C^{-d}$. We now apply Theorem 7 to the degree-$2d$ polynomial $q(x) = -e(x) + \widehat{e}(\emptyset)$, and we see that with probability at least $C^{-d}$ (over a random uniform draw of $x$) we have $-e(x) + \widehat{e}(\emptyset) > C^{-d}$, and hence $\mathbf{Pr}_x[\text{sign}(e(x)) < 0] > C^{-d}$.

We now observe that $\text{sign}(e(x)) < 0$ if and only if $\text{sign}(a(x)) \neq \text{sign}(b(x))$, and consequently $\mathbf{Pr}_x[\text{sign}(e(x)) < 0]$ is precisely $\text{dist}(\text{sign}(a), \text{sign}(b))$. We thus have that for $a(x), b(x)$ drawn from $\mathcal{D}$ as described above, the probability that $\text{dist}(\text{sign}(a), \text{sign}(b))$ is less than $C^{-d}$ is at most $2^{-\alpha_d n^d}$ for some absolute constant $\alpha_d > 0$ (depending only on $d$).

Now let us consider $M = 2^{(\alpha_d/2)n^d}$ many independent draws of polynomials $a_1, a_2, \ldots, a_M$ from $\mathcal{D}$. A union bound over all the $\binom{M}{2} < 2^{\alpha_d n^d}$ pairs $(i,j)$ with $1 \leq i < j \leq M$ gives that with nonzero probability, every $a_i, a_j$ pair satisfies $\text{dist}(\text{sign}(a_i), \text{sign}(a_j)) \geq C^{-d}$. Thus there must be some outcome for the polynomials $a_1, a_2, \ldots, a_M$ such that $\text{dist}(\text{sign}(a_i), \text{sign}(a_j)) \geq C^{-d}$ for all $1 \leq i < j \leq M$. Setting $f_i = \text{sign}(a_i)$ for this outcome, Theorem 10 is proved. $\square$

It remains only for us to prove Lemma 15.

### D.2.2 Proof of Lemma 15

Let us consider $a(x) = \sum_{|S|=d} \widehat{a}(S) \chi_S(x)$ and $b(x) = \sum_{|S|=d} \widehat{b}(S) \chi_S(x)$ drawn independently from $\mathcal{D}$. We will show that the bias of the polynomial $c(x) = a(x)b(x)$ fails to satisfy the bound in item 1 with probability $2^{-\Omega_d(n^d)}$. Then we show the variance of $c$ fails to satisfy item 2 with probability $2^{-\Omega_d(n^d)}$, and the lemma follows from a union bound.

To bound the bias of $c$, we begin by noting that:

$$\widehat{c}(\emptyset) = \sum_{S \subseteq [n]} \widehat{a}(S) \widehat{b}(S).$$

Each term $\widehat{a}(S)\widehat{b}(S)$ in the summand is uniform, i.i.d in $\{-1, 1\}$. Define the random variable $X_S = 1/2 - (1/2)\widehat{a}(S)\widehat{b}(S)$. Then $\sum_{S \subseteq [n]} X_S$ is binomially distributed and setting $t = \frac{1}{4} C^{-d} (\frac{n}{2d})^d$, we may apply the

Chernoff bound to obtain:

$$\mathbf{Pr}[\widehat{c}(\emptyset) < -\frac{1}{4}C^{-d}(\frac{n}{2d})^d] = \mathbf{Pr}[X > \mathbf{E}[X] + t] \leq \exp{(-2\frac{t^2}{\binom{n}{d}})} = 2^{-\Omega_d(n^d)}$$

The same analysis gives a bound on the magnitude in the negative direction. Since $\binom{n/2}{d} \geq (\frac{n}{2d})^d$, this concludes the analysis for the first item of the lemma.

Now we show that item 2 of the lemma also fails with very small probability. The following terminology will be useful. Let $T \subset [n]$ be a subset of size exactly $|T| = 2d$ (we think of $T$ as the set of variables defining some monomial of degree $2d$). For such a $T$ we let $\mathrm{first}(T) \stackrel{\mathrm{def}}{=} T \cap [n/2]$ and $\mathrm{second}(T) \stackrel{\mathrm{def}}{=} T \cap [n/2+1, n]$. We say that such a $T$ is *balanced* if $|\mathrm{first}(T)| = |\mathrm{second}(T)| = d$. Note that there are exactly $\binom{n/2}{d}^2$ many balanced subsets $T$.

We say that a subset $U \subset [n], |U| = d$ is *pure* if $U$ is contained entirely in $[n/2+1, n]$.

Let us consider $a(x) = \sum_{|S|=d} \widehat{a}(S)\chi_S(x)$ and $b(x) = \sum_{|S|=d} \widehat{b}(S)\chi_S(x)$ drawn independently from $\mathcal{D}$. Fix any outcome for $a$ (i.e. for the values of all $\binom{n}{d}$ coefficients $\widehat{a}(S)$), and fix any outcome for $\widehat{b}(U)$ for every $U$ which is *not* pure. Thus the only "remaining randomness" is the value (drawn uniformly from $\{-1, 1\}$) for each of the $\binom{n/2}{d}$ coefficients $\widehat{b}(U)$ for pure $U$. We will show that with probability at least $1 - 2^{-\Omega_d(n^d)}$ over the remaining randomness, at least $\frac{1}{6}\binom{n/2}{d}^2$ of the $\binom{n/2}{d}^2$ many balanced subsets $T$ have $\widehat{c}(T) \neq 0$. Since each value $\widehat{c}(T)$ which is nonzero is at least 1 in magnitude, this suffices to prove the lemma.

Consider any fixed pure subset $U \subset [n], |U| = d$ (for example $U = \{n-d+1, \ldots, n\}$). Let $T$ be a balanced subset of $n$ (so $|T| = 2d$) such that $\mathrm{second}(T)$ equals $U$. (There are precisely $\binom{n/2}{d}$ balanced subsets $T$ with this property; let $\mathcal{T}_U$ denote the collection of all $\binom{n/2}{d}$ of them.) Consider the value $\widehat{c}(T)$: this is

$$\widehat{c}(T) = \sum_{S \subseteq T, |S|=d} \widehat{a}(S)\widehat{b}(T-S).$$

The only "not-yet-fixed" part of the above expression is the single coefficient $\widehat{b}(U)$; everything else has been fixed. Since the coefficient $\widehat{a}(T-U)$ of $\widehat{b}(U)$ is a nonzero integer, there are two possible outcomes for the value of $\widehat{c}(T)$, depending on whether $\widehat{b}(U)$ is set to +1 or -1. These two possible values differ by 2; consequently, there is at most one possible outcome of $\widehat{b}(U)$ that will cause $\widehat{c}(T)$ to be zero. (Note that it may well be the case that no outcome for $\widehat{b}(U)$ would cause $\widehat{c}(T)$ to become zero.)

Let us say that an outcome of $\widehat{b}(U)$ is *pernicious* if it has the following property: at most $\frac{1}{3}\binom{n/2}{d}$ of the $\binom{n/2}{d}$ elements $T \in \mathcal{T}_U$ have $\widehat{c}(T)$ take a nonzero value under that outcome of $\widehat{b}(U)$. (Equivalently, at least $\frac{2}{3}\binom{n/2}{d}$ of the $\binom{n/2}{d}$ elements $T \in \mathcal{T}_U$ have $\widehat{c}(T)$ become zero under that outcome of $\widehat{b}(U)$.) It may be the case that neither outcome in $\{-1, 1\}$ for $\widehat{b}(U)$ is pernicious (e.g. if each outcome makes at least 95% of the $\widehat{c}(T)$ values come out nonzero). It cannot be the case that both outcomes $\{-1, 1\}$ for $\widehat{b}(U)$ are pernicious (for if there were two pernicious outcomes, this would mean that at least $\frac{1}{3}$ of the $\widehat{c}(T)$ values evaluate to 0 under both outcomes for $\widehat{b}(U)$, but it is impossible for any $\widehat{c}(T)$ to evaluate to 0 under two outcomes for $\widehat{b}(U)$). Consequently we have

$$\mathbf{Pr}[\text{the outcome of } \widehat{b}(U) \text{ is pernicious}] \leq 1/2.$$

This is true independently for each of the $\binom{n/2}{d}$ many pure subsets $U$. As a result, a simple analysis gives

$$\mathbf{Pr}[\text{at least 3/4 of the } \binom{n/2}{d} \text{ pure subsets } U \text{ have a pernicious outcome}] \leq 2^{-\Omega_d(n^d)}.$$

Thus we may assume that fewer than $3/4$ of the $\binom{n/2}{d}$ pure subsets $U$ have a pernicious outcome. So at least $\frac{1}{4}\binom{n/2}{d}$ of the pure subsets $U$ are non-pernicious. For each such non-pernicious $U$, more than $\frac{1}{3}\binom{n/2}{d}$ of the $\binom{n/2}{d}$ elements in $\mathcal{T}_U$ have $\widehat{c}(T)$ take a nonzero value. Consequently, at least $\frac{1}{12}\binom{n/2}{d}^2$ many balanced subsets $T$ overall have $\widehat{c}(T) \neq 0$. This proves the lemma. □